\documentclass[aps,prx,twocolumn,groupedaddress,showpacs,dvipdfmx]{revtex4-1}
\usepackage{graphicx}% Include figure files
\usepackage{dcolumn}% Align table columns on decimal point
\usepackage{bm}% bold math
\usepackage{amsmath}
\usepackage{amssymb}
\usepackage{txfonts}
\usepackage{url}
\usepackage{here}
\usepackage[usenames,dvipsnames]{color}
\definecolor{Gray}{gray}{0.0}
\definecolor{lightGray}{gray}{0.35}

\begin{document}
\title{Exotic phase transition and superconductivity in layered titanium-oxypnictides
  implied by a computational phonon analysis}
\author{Kousuke Nakano$^{1}$}
\email{kousuke\_1123@icloud.com}
\author{Kenta Hongo$^{1, 2, 3}$}
\author{Ryo Maezono$^{1}$}
\email{rmaezono@mac.com}
\affiliation{$^{1}$
  School of Information Science, JAIST, Asahidai 1-1, Nomi, Ishikawa
  923-1292, Japan,}
\affiliation{$^{2}$
  JST-PRESTO, 4-1-8 Honcho, Kawaguchi, Saitama 332-0012, Japan,}
\affiliation{$^{3}$
  Center for Materials research by Information Integration (CMI$^2$), 
  National Institute for Materials Science (NIMS),
  1-2-1 Sengen, Tsukuba, Ibaraki 305-0047, Japan
}

\date{\today}
\begin{abstract}
  \noindent
We applied {\it ab initio} phonon analysis to 
layered titanium-oxypnictides, 
Na$_2$Ti$_2Pn_2$O ($Pn$ = As, Sb), and found a clear 
contrast between the cases with lighter/heavier pnictogen 
in comparisons with experiments. 
The result completely explains the experimental structure, 
$C$2/$m$ for $Pn$ = As, within the conventional electron-phonon 
framework, while there arise discrepancies when the pnictogen %arises
gets heavier, being in the same trend for the BaTi$_2Pn_2$O 
($Pn$ = As, Sb, Bi) case. 
The fact implies a systematic dependence on 
pnictogen to tune the mechanism of the phase transition and superconductivity 
from conventional to exotic.

\vspace{2mm}
\noindent
PACS numbers: 71.45.Lr, 74.25.Kc, 74.70.-b

\end{abstract}
\maketitle

%%%%%%%%%%%%%%%%%%%%%%%%%%%%%%
\section{Introduction}
\label{sec:introduction}
%%%%%%%%%%%%%%%%%%%%%%%%%%%%%%
Recently discovered layered titanium-oxypnictides, $A$Ti$_2Pn_2$O 
[$A =$ Na$_2$, Ba, (SrF)$_2$, (SmO)$_2$; $Pn =$ As, Sb, Bi]
~\cite{1997AXE,2001OZA,2009LIU,2010WAN,2012DOA,2012YAJ,2013YAJ1,2013YAJ2,2013ZHA,2013NAK,2014PAC,2014ROH}, 
is a new superconducting family, 
and their superconducting mechanisms have attracted 
intensive attentions because their crystal and electronic 
structures are similar to the exotic superconductors 
such as cuprates~{\cite{1986BED}} or iron arsenides~{\cite{2008KAM}}.
The mechanism has been expected to be {\it conventional}, 
namely electron-phonon driven 
\cite{2013SUB,2013ROH,2013KIT,2013NOZ}, 
and accompanying singularities in resistivity and susceptibility 
at low temperature are regarded due to 
the conventional charge density wave (CDW) 
{\cite{2013KIT,2013ROH,2013NOZ,2013MEL,2015TAN}}.
Recent works are, however, pointing out 
the possibility of {\it exotic} mechanism 
for these compounds: 
For BaTi$_2Pn_2$O ($Pn$ = As, Sb), 
Frandsen~{\it et al.} reported the breaking four-fold symmetry 
at low temperature without any superlattice peaks by 
Neutron diffraction and TEM{~\cite{2014FRA}}, and suggested the possibility 
of the intra-unit-cell nematic CDW{~\cite{2010LAW,2014FUJ}} 
to make account for the breaking. 
A theoretical suggestion was made 
that such an intra-unit-cell nematic CDW 
could be realized by the orbital ordering 
mediated by spin-fluctuations{~\cite{2016NAK1}}. 
The Uemura classification scheme{~\cite{1991UEM}} 
also supports exotic superconducting mechanisms in these 
compounds{~\cite{2014KAM}}. 
Though our latest phonon evaluations{~\cite{2016NAK}} 
partly refuted that the breaking could be explained 
within the conventional electron-phonon mechanism 
at least for $Pn$ = As, it is still unknown if this is 
the case for $Pn$ = Sb and Bi, leaving debates 
about whether the mechanism is conventional or exotic. 

%%%%%%%%%%%%%%%%%%%%%%%%%%%%%%%%%%%%%
\begin{figure}[htbp]
  \centering
  \includegraphics[width=3.5cm]{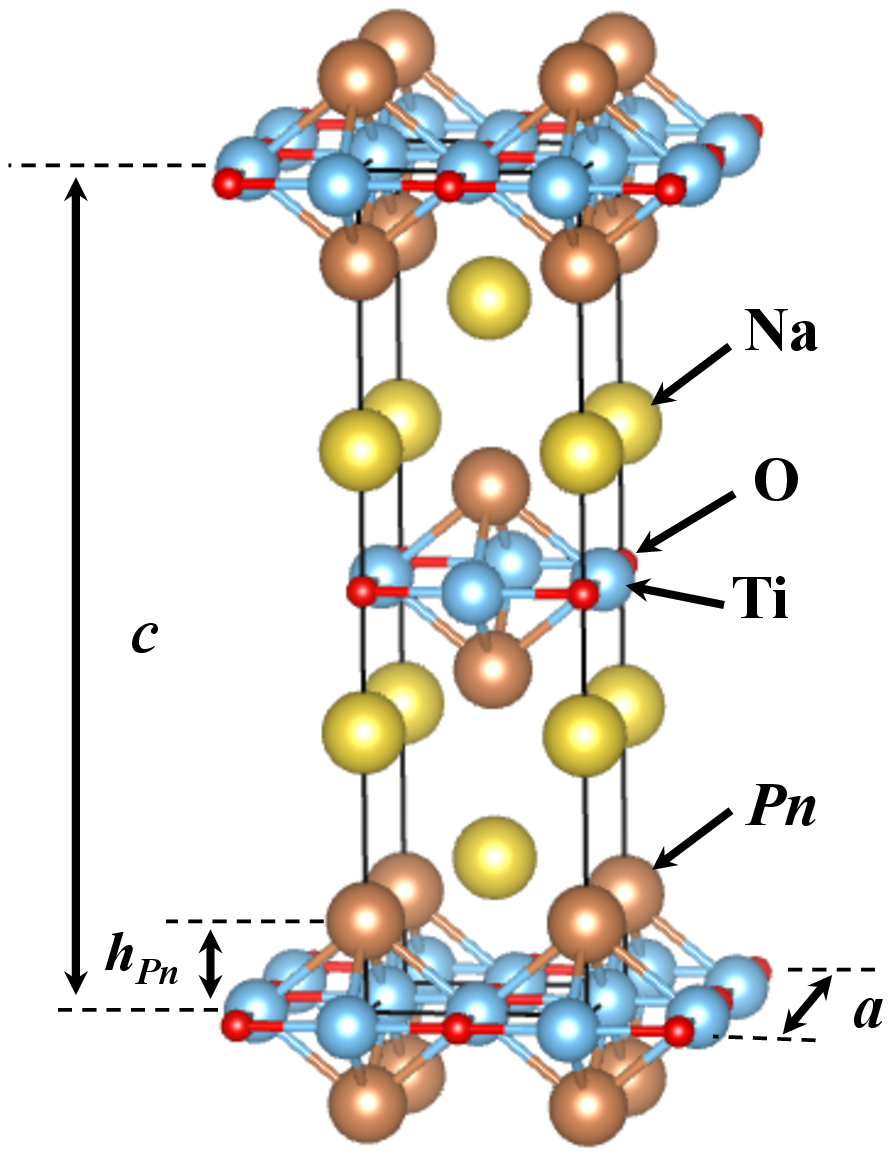}
  \caption{The crystal structure of Na$_2$Ti$_2Pn_2$O ($Pn$ = As, Sb). 
  The space group is $I$4/$mmm$.
  }
  \label{crystal}
\end{figure}
%%%%%%%%%%%%%%%%%%%%%%%%%%%%%%%%%%%%%

\vspace{2mm}
Examining whether the observed superlattices are 
explained by the conventional {\it ab initio} 
framework or not can provide a critical clue 
for the questions. 
For experimental side, Davies~{\it et al.}{~\cite{2016DAV}} 
have just synthesized large single crystals 
of Na$_2$Ti$_2Pn_2$O ($Pn$ = As, Sb) (Fig.~{\ref{crystal}}) in order 
to achieve reliable diffaction measurements, 
and observed clear superlattice peaks of 
$2 \times 2 \times 2$ and $2 \times 2 \times 1$ for $Pn$ = As and Sb,
respectively. 
For $Pn$ = As, a recent phonon calculation\cite{2016CHE} 
reported possible superlattices but they could not
explain the observed peaks. 
In this paper, we report that our 
{\it ab initio} phonon calculations 
can explain the observed $C$2/$m$ (monoclinic No.12) 
structure for $Pn$ = As, implying that 
the conventional electron-phonon driven mechanism 
is likely for this system. 
For $Pn$ = Sb, on the other hand, we obtained 
$Cmce$ (Orthorhombic No.64) superlattice, 
being inconsistent with the observed one 
[$Cmcm$ (Orthorhombic No.63)]. 
The discrepancy would support exotic mechanisms 
likely for $Pn$ = Sb in contrast. 
Based on the results combined with preceding 
studies{\cite{2009KUR,2010MIY,2012SIN,2013SUB,2013YAN,2016NAK}}, 
we propose a new possibility that a heavier $Pn$ causes 
stronger electron correlations and 
induces exotic phase transition and superconductivity that cannot be 
explained by the conventional framework.

%%%%%%%%%%%%%%%%%%%%%%%%%%%%%%%%%%%%
\begin{figure*}[htbp]
  \centering
  \includegraphics[width=8.0cm]{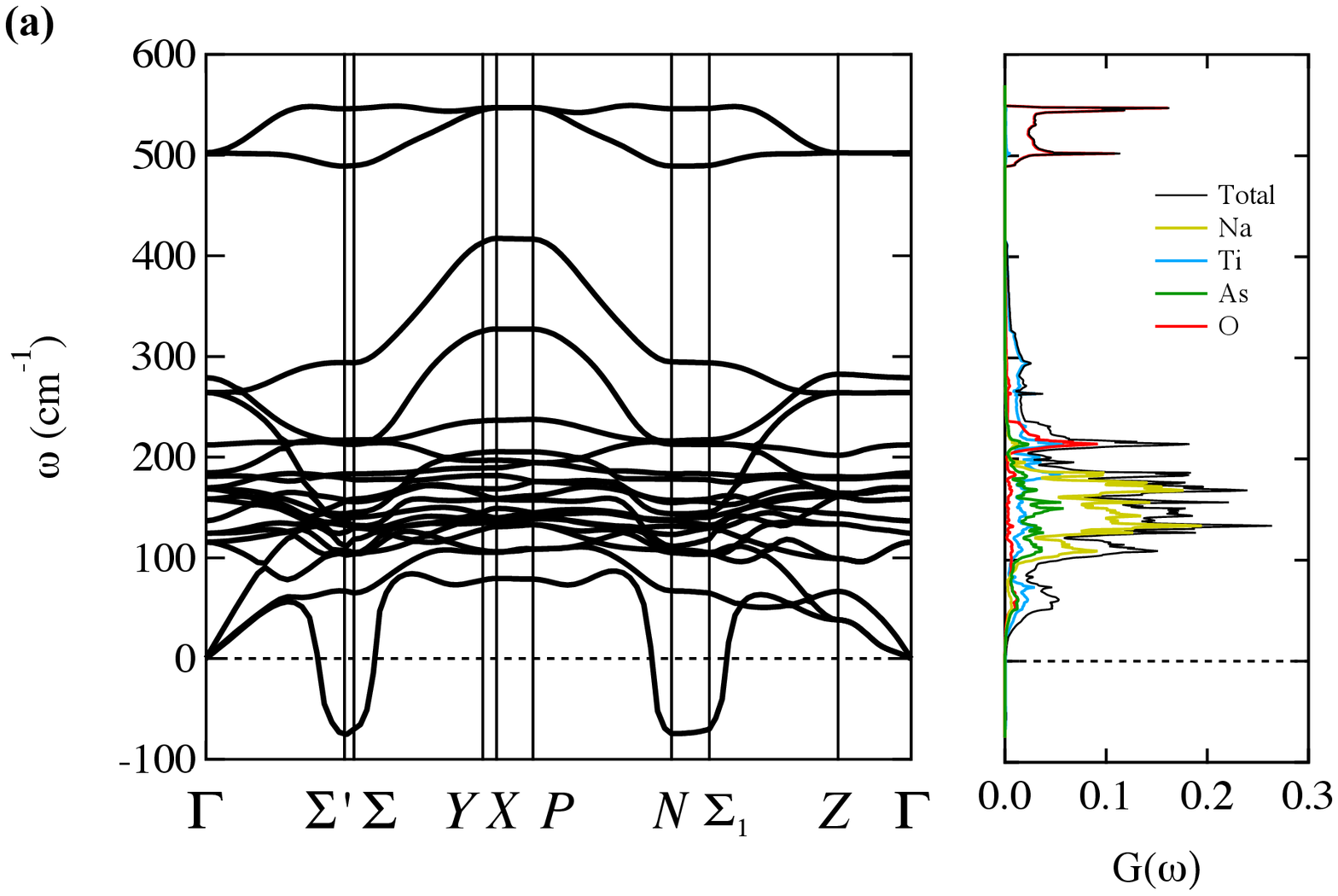}
  \includegraphics[width=8.0cm]{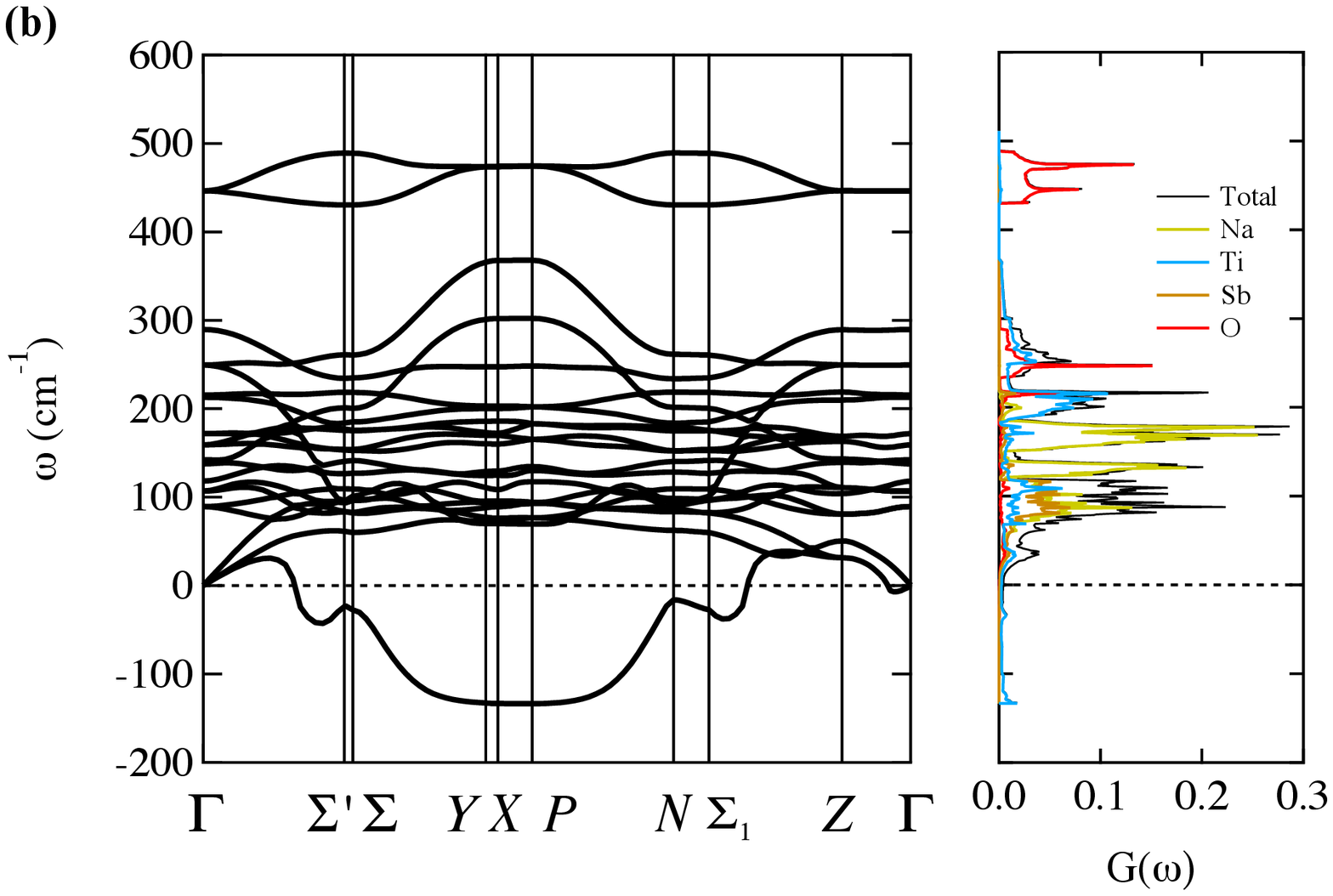}
  \caption{Phonon dispersions and phonon DOS for undistorted ($I$4/$mmm$)     
  Na$_2$Ti$_2Pn_2$O ($Pn$ = (a) As, (b) Sb).
  The primitive Brillouin zones are shown in Fig.~{\ref{SI_BZ}} (a)}.
  \label{ph_ntao}
\end{figure*}
%%%%%%%%%%%%%%%%%%%%%%%%%%%%%%%%%%%%

%%%%%%%%%%%%%%%%%%%%%%%%%%%%%%%
\begin{figure}[htbp]
  \centering
  \includegraphics[width=6.5cm]{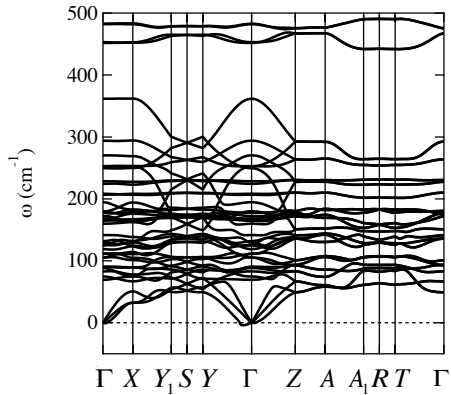}
  \caption{Phonon dispersions for Na$_2$Ti$_2$Sb$_2$O superlattice ($Cmce$) .
  The primitive Brillouin zone is shown in Fig.~{\ref{SI_BZ} (b)}.
  }
  \label{ph_2_2_1_ntso}
\end{figure}

%%%%%%%%%%%%%%%%%%%%%%%%%%%%%%%

%%%%%%%%%%%%%%%%%%%%%%%%%%%%%%%%%%%%%
\section{Method}
\label{sec:method}
All the calculations were done within DFT 
using GGA-PBE exchange-correlation functionals~\cite{1996PER}, 
implemented in Quantum Espresso package.~\cite{2009PAO}
We adopted PAW~\cite{1994BLO} pseudo potentials. 
The present PAW implementation
takes into account the scalar-relativistic effects 
upon a careful comparison with all-electron calculations.~{\cite{2014KUC1,2014KUC2}}
We restricted ourselves to spin unpolarized calculations.
Lattice instabilities were detected by the negative (imaginary) 
phonon dispersions evaluated for undistorted and distorted structures.
Taking each of the negative phonon modes, the structural relaxations along
the mode were evaluated by the BFGS optimization scheme with the 
structural symmetries fixed to $C$2/$m$ for Na$_2$Ti$_2$As$_2$O,
$Cmce$ for Na$_2$Ti$_2$Sb$_2$O. 
For phonon calculations, we used the linear response theory implemented in
Quantum Espresso package.~{\cite{2001BAR}}
Crystal structures and Fermi surfaces were depicted by using 
VESTA~\cite{2011MOM} and XCrySDen~\cite{1999KOK}, respectively.

\vspace{2mm}
To deal with all the compounds systematically, we checked the convergence of
plane-wave cutoff energies ($E_{\rm cut}$), $k$-meshes, $q$-meshes,
and smearing parameters.
The most strict condition among the compounds was taken 
to achieve the convergence within $\pm 1.0$ mRy per formula unit 
in the ground state energy, 
resulting in $E_{\rm cut}^{(\rm WF)} = 90~ $ Ry for wavefunction and
$E_{\rm cut}^{(\rho)} = 800~ $ Ry for charge density.
For undistorted Na$_2$Ti$_2$As$_2$O and Na$_2$Ti$_2$Sb$_2$O, 
($6 \times 6 \times 6$) $k$-meshes were used for the
Brillouin-zone integration. 
Phonon dispersions were calculated on ($6 \times 6 \times 6$) $q$-meshes.
For distorted Na$_2$Ti$_2$As$_2$O ($C$2/$m$) and Na$_2$Ti$_2$Sb$_2$O ($Cmce$) 
superlattices, ($4 \times 4 \times 4$) $k$-meshes were used.
For the distorted Na$_2$Ti$_2$Sb$_2$O superlattice, ($4 \times 4 \times 4$)
$q$-meshes were used to calculate phonon dispersions.
We also calculated a ground state energy of the experimentally 
observed Na$_2$Ti$_2$Sb$_2$O ($Cmcm$) superlattice~{\cite{2016DAV}}
with ($6 \times 6 \times 3$) $k$-meshes.
The Marzari-Vanderbilt cold smearing scheme~\cite{1999MAR} with
a broadening width of 0.01 Ry was applied to all the compounds.
All the calculations were performed with the primitive lattices.

%%%%%%%%%%%%%%%%%%%%%%%%%%%%%%%%%%%%
\begin{figure*}[htbp]
  \centering
  \includegraphics[width=15cm]{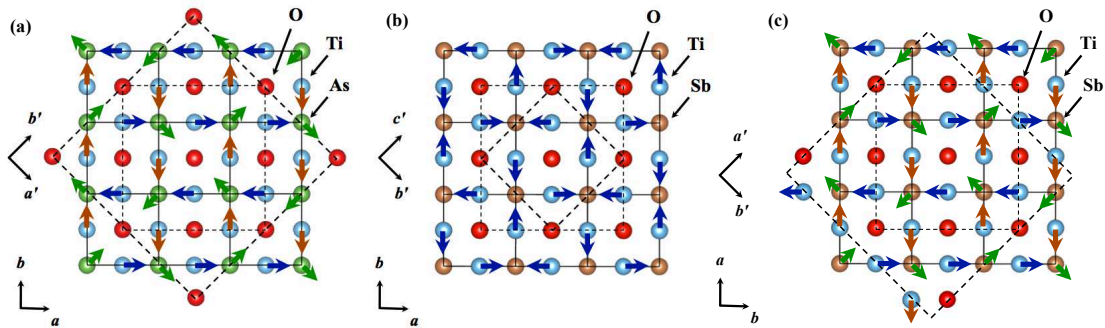}
  \caption{In-plane superlattice structures obtained by our calculations 
  for $Pn$ = (a) As and (b) Sb, and (c) the observed in-plane superlattice structure 
  for $Pn$ = Sb{~\cite{2016DAV}}.
  $Pn$ is located above and below the Ti$_2$O plane.
  Solid lines represent the original unit cells for $I$4/$mmm$. 
  Thin dash lines represent the unit cells for $2 \times 2$ superlattices.
  Solid dash lines represent redefined unit cells for distorted superlattices.
  }
  
  \label{dis_2D}
\end{figure*}
%%%%%%%%%%%%%%%%%%%%%%%%%%%%%%%

%%%%%%%%%%%%%%%%%%%%%%%%%%%%%%%
\begin{figure}[htbp]
  \centering
  \includegraphics[width=8.0cm]{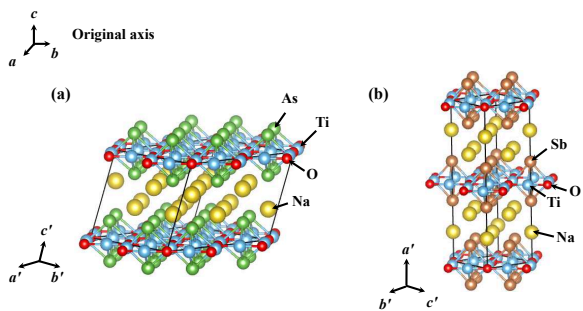}
  \caption{3D superlattice structures obtained from our calculations for 
  $Pn$ = (a) As ($C$2/$m$) and (b) Sb ($Cmce$).
  }
  \label{dis_3D}
\end{figure}
%%%%%%%%%%%%%%%%%%%%%%%%%%%%%%%

%%%%%%%%%%%%%%%%%%%%%%%%%%%%%%%%%%
\section{Results and Discussion}
\label{sec:results}
%%%%%%%%%%%%%%%%%%
Figure~{\ref{ph_ntao}} (a) shows the phonon dispersions
for $Pn$ = As, giving imaginary frequencies 
appearing around $N$ = $(2\pi/a)\cdot$(1/2, 0, $a/c$) 
and $\Sigma'$ = $(2\pi/a)\cdot$(1/2, 0, 0),
where ${a}$ and ${c}$ are conventional 
lattice constants for the undistorted structure. 
The compatible modes with the experimental 
observation{~\cite{2016DAV}} that 
the twice larger periodicity along $c$-axis 
realized are 
$N$ = $(2\pi/a)\cdot$(1/2, 0, $a/c$) and 
$N'$ = $(2\pi/a)\cdot$(0, 1/2, $a/c$). 
We note that the previous phonon calculation{~\cite{2016CHE}} 
could not reproduce this doubled periodicity. 
We therefore took these for 
further lattice relaxations from 
the original $I$4/$mmm$ structure 
along the mode displacements to get 
$C$2/$m$ (monoclinic, No.12) superlattice 
(Fig.~{\ref{dis_2D}} (a) and Fig.~{\ref{dis_3D}} (a)), 
being consistent with the experiments{~\cite{2016DAV}}.
The resultant optimized geometry after the relaxation 
gives fairly good agreement
with the experiments{~\cite{2016DAV}}
within deviations at most 2.5\% (See Appendix C).
We could not confirm whether or not the negative modes 
disappear in the $C$2/$m$ superlattice 
just because of intractable computational costs 
for enlarged reciprocal space by the lowered symmetry. 

\vspace{2mm}
For $Pn$ = Sb, we obtained phonon dispersions as 
shown in Fig.~\ref{ph_ntao} (b).
Imaginary frequencies appear widely around 
$X$ = $(2\pi/a)\cdot$(1/2, 1/2, 0) 
and $P$ = $(2\pi/a)\cdot$(1/2, 1/2, $a/2c$). 
The experiments{~\cite{2016DAV}} reported that
the twice larger super periodicity is 
realized only in $ab$-plane, not along $c$-axis. 
It is therefore $X$ mode being likely 
to realize the instability toward the superlattice. 
By optimizing the geometry along the mode displacement, 
we obtained $Cmce$ (orthorhombic, No.64) 
(Fig.~{\ref{dis_2D}} (b) and Fig.~{\ref{dis_3D}} (b)), 
being inconsistent with the observed $Cmcm$ (orthorhombic, No.63) 
structure (Fig.~\ref{dis_2D} (c)) in the experiment{\cite{2016DAV}}. 
To confirm the present conclusion further, 
we examined the stability by seeing if 
the imaginary frequencies disappear 
at the relaxed structure, as shown 
in Fig.~\ref{ph_2_2_1_ntso}. 
In addition to confirming the stability, 
we compared enthalpies between our $Cmce$ 
and the observed $Cmcm$. 
The optimized geometry with $Cmce$ is found to be
more stable by 11 mRy per formula unit than that with $Cmcm$.
These results mean that the experimentally observed $Cmcm$ 
is energetically unfavorable within DFT framework.

\vspace{2mm}
Putting the discrepancy for $Pn$ = Sb aside for a while, 
we can provide a plausible explanation for the 
lattice instabilities as follows: 
The lattice instability 
mainly derives from in-plane Ti vibration.
To make account for the inter-atomic forces 
to restore the lattice displacements, we can 
ignore (at least for the discussion of 
the instability) the contribution from Na
because there are no bonds between Na and Ti (Fig.~{\ref{crystal}}).
Starting with $X$ mode for $Pn$ = Sb (Fig.~\ref{dis_2D} (b)), 
it generates a displacement such that 
a rectangular formed by Ti 
rotates to vibrate within $ab$-plane.
A possible restoring force would come from 
inter-ion interactions between Ti$^{3+}$ and
nearest O$^{2-}$, but the force is orthogonal 
to the displacement and hence we can ignore it. 
Primary force would therefore come from 
Ti-$Pn$ interaction for Ti approaching to $Pn$ 
when the rectangular rotates. 
The trend in the electronegativity 
(As = 2.18, Sb = 2.05) implies a weaker Ti-$Pn$
interaction for $Pn$ = Sb. 
The fact that we get the instability only for 
$Pn$ = Sb implies that the threshold would be 
around 2.1 in terms of the electronegativity 
against the instability. 
For $N$ mode (Fig.~\ref{dis_2D} (a)), the imaginary frequency appears 
commonly for $Pn$ = As and Sb. 
This is also accountable along the above discussion. 
In this mode, the Ti-$Pn$ interaction has litte effect 
on the restoring forces because Ti and $Pn$ move in 
the same direction.
The possible restoring force would originate from 
nearest Oxygen and again this is orthogonal 
to the displacement, giving too weak contribution 
leading to the instability. 
This is the reason why the imaginary frequencies appear 
at $N$ mode for both $Pn$.

\vspace{2mm}
Getting back to the discrepancy, 
the present results confront 
the sharp contrast between $Pn$ = As and Sb 
in terms of whether the conventional electron-phonon
treatment for the phase transition works or not.
Since it is clearly reported{~\cite {2016DAV}} 
that the measured XRD results cannot be identified 
by the $\sqrt 2\times\sqrt 2\times 1$ superlattice (Fig.~\ref{dis_2D} (b)), 
the discrepancy would be intrinsic so that 
it should be accounted for by those mechanism 
beyond the conventional ones, such as 
strong electronic correlations. 
As discussed above, the present predictions 
about the instabilities even including $Pn$ = Sb 
can be explained to some extent 
by a plausible physical picture, 
but another mechanism could dominate over this 
to critically decide which mode is chosen 
as the phonon condensation mode when the temperature 
decreases.

\vspace{2mm}
One of the possible mechanisms would be 
the spatial anisotropy introduced via 
spin-orbit couplings{~\cite{2015KIM}} under the enhanced 
polarizations by the electronic 
correlation: \cite{2009KUR,2010MIY}
In layered titanium-oxypnictides,
the hybridization between Ti-3$d$ and 
$p$-orbitals of $Pn$
{\cite{2012SIN,2013SUB,2013YAN,2016NAK}}
is one of the critical factor for 
the transport property about how much 
the valence electrons are localized. 
For iron arsenide superconductors, 
such tendency toward the localization is 
well captured by the trend of $h$, 
the vertical distances between Fe layer and 
$Pn$ or $Ch${~\cite{2009KUR,2010MIY}} (Fig.~\ref{crystal}).
When $h$ gets larger, the covalency gets 
weaker to make Fe-3$d$ more localized 
and then the spin/orbital polarizations get 
enhanced as one of the electronic correlation 
effect.{~\cite{2009KUR,2010MIY}}
If we apply the similar analysis to
the present case, we get 
$h_{Pn = Bi} > h_{Pn = Sb} > h_{Pn = As}$ 
{\cite{2013YAJ1,2012YAJ,2012DOA,2000OZA,2010LIU,2010WAN}} 
for the experimental geometry. 
This trend would support the correlation effect 
gets more enhanced for Sb than As.

\vspace{2mm}
The trend in $h$ above is again 
consistent with our previous study 
for BaTi$_2Pn_2$O ($Pn$ = As, Sb, Bi), where 
we showed that the conventional electron-phonon framework 
could explain experiments only for $Pn$ = As, but not for the other
heavier $Pn$.~{\cite{2016NAK}} 
It might be a general tendency also applicable 
to the layered titanium-oxypnictides that the larger $h$ 
enhances the electronic correlations as the origin 
of the exotic mechanism. 
Such mechanisms have actually been proposed 
by several authors{~\cite{2015KIM,2016NAK1}} such as 
the phase transition driven not by CDW but by the 
orbital ordering{~\cite{2015KIM}}, 
or the orbital ordering induced by 
spin-fluctuation{~\cite{2016NAK1}}. 
We note that there are some papers 
reporting 'weak correlations' 
in these compounds{~\cite{2013YAN,2013HUA,2015KIM,2015TAN}}. 
It apparently seems contrary
but we have to be careful that 
some reports support 'weaker than Fe-based compounds' 
\cite{2013HUA} 
while other 'weaker correlation in XC potentials'. 
\cite{2015TAN,2013YAN}
These are therefore not necessarily 
contrary to 
the present statement that 
'the correlation gets stronger to overwhelm  
the conventional electron-phonon mechansm'. 
The above exotic orbital orderings{~\cite{2015KIM,2016NAK1}} 
are reported to be possible even when 'the correlation 
is weak', but again we wonder if this would be such a 
situation as 
$E_{\rm corr.}^{\rm Fe} > E_{\rm corr.}^{\rm Ti} > E_{ph.}$, 
where $E_{\rm corr.}$ and $E_{ph.}$ correspond to 
the typical energy scale of electronic correlations 
and electron-phonon interaction, respectively. 
Though the superconductivities in BaTi$_2$$Pn_2$O ($Pn$ = Sb, Bi) 
have been thought to be conventional 
BCS type{~\cite{2013SUB,2013KIT,2013NOZ,2013ROH}}, 
the present result implies the possibility of 
some exotic superconducting mechanism with heavier $Pn$. 
We also note that it was pointed out{~\cite{2014KAM}} that 
the superconductivity in Ba$_{1-x}$Na$_x$Ti$_2$Sb$_2$O 
is at the verge of unconventional superconductivity by 
the Uemura classification scheme{~\cite{1991UEM}}.

%%%%%%%%%%%%%%%%%%%%%%%%%%%%%%%
\section{Conclusion}
\label{sec:conclusion}
%%%%%%%%%%%%%%%%%%%%%%%%%%%%%%%
We performed {\it ab-initio} phonon calculations for Na$_2$Ti$_2Pn_2$O ($Pn$ = As, Sb) 
to investigate if the experimentally observed superlattice peaks can be explained 
by negative frequency modes.
For $Pn$ = As, we obtained a $C$2/$m$ (monoclinic No.12) superlattice
that is completely consistent with the experimentally observed structure. 
The consistency indicates that simple electron-phonon interaction can explain 
the phase transition for $Pn$ = As. 
On the other hand, we obtained a $Cmce$ (orthorhombic No.64) 
superlattice for $Pn$ = Sb that is inconsistent with the experimentally observed 
structure $Cmcm$ (orthorhombic No.63). 
The discrepancy would be intrinsic so that it should be accounted for by those mechanism 
beyond electron-phonon interaction, such as strong electronic correlations. 
Such a discrepancy is also found in superconducting BaTi$_2Pn_2$O 
when the pnictogen gets heavier.
These facts imply a systematic dependence on 
pnictogen to tune the mechanism of the phase transition and superconductivity
from conventional to exotic.

%%%%%%%%%%%%%%%%%%%%%%%%%%%%%%%
\begin{acknowledgments}
\end{acknowledgments}
The computation in this work has been mainly
performed using the facilities of the Center for Information Science in JAIST.
K.H. is grateful for financial support from a KAKENHI grant (15K21023), a Grant-in-Aid for Scientific Research on Innovative Areas (16H06439), PRESTO and the Materials research by Information Integration Initiative (MI$^2$I) project of the Support Program for Starting Up Innovation Hub from Japan Science and Technology Agency (JST).
R.M. is grateful for financial support from MEXT-KAKENHI 
grants 26287063 and that from the Asahi glass Foundation. 

%%%%%%%%%%%%%%%%%%%%%%%%%%%%%%%
%APPENDIX
%%%%%%%%%%%%%%%%%%%%%%%%%%%%%%%

\renewcommand{\refname}{}
\makeatletter
\makeatother

%%%%%%%%%%%%%%%%%%%%%%%%%%%%%%%
\section*{Appendix A : Electronic structures}
\label{Appendix_A}
%%%%%%%%%%%%%%%%%%%%%%%%%%%%%%%

\begin{figure}[htbp]
  \centering
  \includegraphics[width=8.0cm]{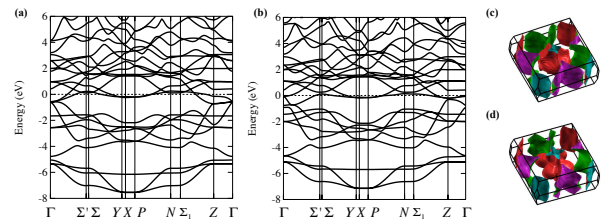}
  \caption{Band structures and Fermi surfaces for undistorted Na$_2$Ti$_2Pn_2$O 
  structure ((a) and (c) $Pn$ = As, (b) and (d) $Pn$ = Sb)}
  \label{SI_1_1_1_Bands_and_FS}
\end{figure}

%% \vspace{2mm}
We described electronic structures for undistorted 
Na$_2$Ti$_2$$Pn_2$O ($Pn$ = As, Sb) structure 
in order to carefully examine the artifacts due to 
the choice of pseudo potentials (PP). 
The obtained electronic structures shown in Fig.~{\ref {SI_1_1_1_Bands_and_FS}}
are consistent with the previous calculations{~\cite{2013SUE2,2013YAN}}.

%%%%%%%%%%%%%%%%%%%%%%%%%%%%%%%
%%%%%%%%%%%%%%%%%%%%%%%%%%%%%%%

%%%%%%%%%%%%%%%%%%%%%%%%%%%%%%%
\section*{Appendix B : Brillouin zones}
\label{Appendix_B}
%%%%%%%%%%%%%%%%%%%%%%%%%%%%%%%

\begin{figure}[htbp]
  \centering
  \includegraphics[width=8.0cm]{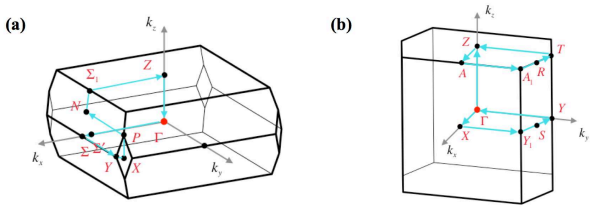}
  \caption{Primitive Brillouin zones (a) for undistorted Na$_2$Ti$_2Pn_2$O structure ($I$4/$mmm$) and
  (b) for Na$_2$Ti$_2$Sb$_2$O superlattice ($Cmce$).}
  \label{SI_BZ}
\end{figure}

%% \vspace{2mm}
The primitive Brillouin zone for undistorted Na$_2$Ti$_2Pn_2$O ($I$4/$mmm$) is
shown in Fig.~{\ref{SI_BZ}} (a).
The special $k$ and $q$ points are
${\Gamma}$ = $(2\pi/a)\cdot$(0, 0, 0), 
$N$ = $(2\pi/a)\cdot$(1/2, 0, $a/2c$), 
$P$ = $(2\pi/a)\cdot$(1/2, 1/2, $a/2c$), 
$X$ = $(2\pi/a)\cdot$(1/2, 1/2, 0), 
$\Sigma'$ = $(2\pi/a)\cdot$(1/2, 0, 0),
$Z$ = $(2\pi/a)\cdot$(0, 0, $a/c$)
in cartesian axis, 
where ${a}$ and ${c}$ are conventional 
lattice constants for the undistorted structures.
%%%%%%%%%%%%%%%%%%%%%%%%%%%%%%%
%%%%%%%%%%%%%%%%%%%%%%%%%%%%%%%
The primitive Brillouin zone for Na$_2$Ti$_2$Sb$_2$O superlattice ($Cmce$) 
is also shown in Fig.~{\ref{SI_BZ}} (b).
The special $k$ and $q$ points are 
${\Gamma}$ = $(2\pi/a')\cdot$(0, 0, 0), 
$X$ = $(2\pi/a')\cdot$(1, 0, 0), 
$S$ = $(2\pi/a')\cdot$(1/2, $a'/2b'$, 0),
$Z$ = $(2\pi/a')\cdot$(0, 0, $a'/2c'$), 
$A$ = $(2\pi/a')\cdot$(1, 0, $a'/2c'$),
$R$ = $(2\pi/a')\cdot$(1/2, $a'/2b'$, $a'/2c'$)
in cartesian axis, 
where ${a'}$, ${b'}$ and ${c'}$ are conventional 
lattice constants for the superlattice structure.

%%%%%%%%%%%%%%%%%%%%%%%%%%%%%%%
\section*{Appendix C : Geometry optimizations}
\label{Appendix_C}
%%%%%%%%%%%%%%%%%%%%%%%%%%%%%%%

%% \vspace{2mm}
The results of geometry optimizations are shown 
in Table {\ref{undistorted_str}}-{\ref{exp_super_str}}.
The conventional lattice vectors of the superlattice structure 
for $Pn$ = As ($C$2/$m$) are redefined as
${\vec a}' = 2\left( {\vec a} - {\vec b} \right)$, 
${\vec b}' = 2\left( {\vec a} + {\vec b} \right)$ and
${\vec c}' = 1/2\left( -{\vec a} + {\vec b} + {\vec c} \right)$,
those for $Pn$ = Sb ($Cmce$) are redefined as
${\vec a}' = {\vec c}$, ${\vec b}' = \left( {\vec a} - {\vec b} \right)$ and 
${\vec c}' = \left( {\vec a} + {\vec b} \right)$, 
and those for $Pn$ = Sb ($Cmcm$) are redefined as 
${\vec a}' = 2\left( {\vec a} + {\vec b} \right)$, 
${\vec b}' = 2\left( {\vec b} - {\vec a} \right)$ and
${\vec c}' = {\vec c}$,
where ${\vec a}$, ${\vec b}$ and ${\vec c}$ are conventional 
lattice vectors of the undistorted structures.

%%%%%%%%%%%%%%%%%%%%%%%%%%%%%%%
\begin{table}[htbp]
  \caption{The results of geometry optimization for undistorted Na$_2$Ti$_2Pn_2$O structure ($Pn$ = As, Sb).
  The relaxed conventional lattice parameters are $a$ = 4.071128 \AA, $c$ = 15.440517 \AA, 
  ($a$ = 4.13957 \AA, $c$ = 16.97766 \AA) for $Pn$ = As (Sb).
  The final enthalpy is -980.0514 Ry (-1323.3103 Ry) 
  for $Pn$ = As (Sb) per formula unit [Na$_2$Ti$_2Pn_2$O].}

\label{undistorted_str}  
\begin{center}
  \begin{tabular}{ccccccc}
       & $Pn$ = As ($I$4/$mmm$)      &     &     & & \\
\hline & label & $x$ & $y$ & $z$ & wyckoff \\
\hline
 & Na &  0.50000  &  0.50000  &  0.18146  &  4$e$   &  \\
 & Ti &  0.50000  &  0.00000  &  0.00000  &  4$c$   &  \\
 & As &  0.00000  &  0.00000  &  0.11752  &  4$e$   &  \\
 &  O &  0.50000  &  0.50000  &  0.00000  &  2$b$   &  \\
\hline
\\
       & $Pn$ = Sb ($I$4/$mmm$)     &     &     & & \\
\hline
 & Na &  0.50000  &  0.50000  &  0.18369  &  4$e$   &  \\
 & Ti &  0.50000  &  0.00000  &  0.00000  &  4$c$   &  \\
 & Sb &  0.00000  &  0.00000  &  0.12015  &  4$e$   &  \\
 &  O &  0.50000  &  0.50000  &  0.00000  &  2$b$   &  \\
\hline
\end{tabular}
\end{center}
\end{table}
%%%%%%%%%%%%%%%%%%

For $Pn$ = As ($C$2/$m$), we need to take special care 
when comparing our calculated geometry with an experimental one 
due to Davis et al.[28] 
We found that the atomic positions listed in their Table I 
are incompatible with their superlattice vectors, 
${\vec a}' = 2\left({\vec a} + {\vec b}\right)$, 
${\vec b}' = 2\left({\vec b} - {\vec a}\right)$, and 
${\vec c}' = 1/2 \left({\vec a} + {\vec b} + {\vec c}\right)$
, described in their main text [28]: 
(1) If the description in the main text is assumed to be correct, 
Na and Sb positions in the superlattice are 
respectively located unlikely far from their undistorted positions. 
(2) O is not located at (0.0,0.0) in their Fig.~6 (a), 
being inconsistent with their Table I. 
Accordingly, we speculate that they again define the superlattice vectors 
compatible with the atomic positions in their Table I, 
which differs from that in their main text.
We also found $y$ = 0.0 for As3 in their Table I must be a typo and correctly
$y$ = 0.25 because As3 should occupy 8$j$ Wyckoff site and its undistorted position is $y$ = 0.5.  
Assuming our speculation is correct, our optimized geometry parameters 
agree well with their experimental values.

%% %%%%%%%%%%%%%%%
\begin{table}[htbp]
\caption{The results of geometry optimization for 
Na$_2$Ti$_2Pn_2$O superlattice ($Pn$ = As, Sb) obtained by phonon calculations.
The conventional lattice parameters after relaxation are $a'$=11.51310 \AA, $b'$ = 11.51244 \AA, 
$c'$ = 8.23431 \AA, $\beta$ = 110.45774$^\circ$ 
($a'$ = 17.04343 \AA, $b'$ = 5.83920 \AA, $c'$ = 5.83039 \AA) for $Pn$ = As (Sb).
The final enthalpy is -980.0517 Ry (-1323.3120 Ry) for $Pn$ = As (Sb) 
per formula unit [Na$_2$Ti$_2Pn_2$O].
}

\label{calc_super_str}  
\begin{center}
  \begin{tabular}{ccccccc}
       & $Pn$ = As ($C$2/$m$)      &     &     & & \\
\hline & label & $x$ & $y$ & $z$ & wyckoff \\
\hline
 & Na &  0.90907  &  0.00000  &  0.63699  &  4$i$   &  \\
 & Na &  0.40943  &  0.00000  &  0.63694  &  4$i$   &  \\
 & Na &  0.15932  &  0.24981  &  0.63725  &  8$j$   &  \\
 & Ti &  0.13031  &  0.11979  &  0.00000  &  8$j$   &  \\
 & Ti &  0.88019  &  0.36969  &  0.00000  &  8$j$   &  \\
 & As &  0.80877  &  0.00000  &  0.23625  &  4$i$   &  \\
 & As &  0.30933  &  0.00000  &  0.23625  &  4$i$   &  \\
 & As &  0.05905  &  0.24969  &  0.23623  &  8$j$   &  \\
 &  O &  0.00000  &  0.00000  &  0.00000  &  2$a$   &  \\
 &  O &  0.25000  &  0.25000  &  0.00000  &  4$e$   &  \\
 &  O &  0.00000  &  0.50000  &  0.00000  &  2$b$   &  \\
 
\hline
\\
       & $Pn$ = Sb ($Cmce$)      &     &     & & \\
\hline
 & Na  &  0.18411   &  0.00000    &  0.00000  & 8$d$   &  \\
 & Ti  &  0.00000   &  0.23266    &  0.26812  & 8$f$   &  \\
 & Sb  &  0.37883   &  0.00000    &  0.00000  & 8$d$   &  \\
 &  O  &  0.00000   &  0.00000    &  0.00000  & 4$a$   &  \\
 
\hline
\end{tabular}
\end{center}
\end{table}
%%%%%%%%%%%%%%%%%%

%% %%%%%%%%%%%%%%%
\begin{table}[htbp]
\caption{The results of geometry optimization for experimentally observed 
Na$_2$Ti$_2$Sb$_2$O superlattice.
The conventional lattice parameters after relaxation are 
$a'$ = 11.73782 \AA, $b'$ = 11.73541 \AA, and $c'$ = 16.82525 \AA.
The final enthalpy is -1323.3109 Ry per formula unit [Na$_2$Ti$_2$Sb$_2$O].
}
\label{exp_super_str}  
\begin{center}
  \begin{tabular}{ccccccc}
       & $Pn$ = Sb ($Cmcm$)      &     &     & & \\
\hline & label & $x$ & $y$ & $z$ & wyckoff \\
\hline
 & Na  &  0.00000   &  0.37498    &  0.93293  & 8$f$   &  \\
 & Na  &  0.00000   &  0.12503    &  0.43294  & 8$f$   &  \\
 & Na  &  0.74999   &  0.12500    &  0.93275  & 16$h$   &  \\
 & Ti  &  0.37417   &  0.00083    &  0.25000  & 8$g$   &  \\
 & Ti  &  0.62584   &  0.24918    &  0.25000  & 8$g$   &  \\
 & Ti  &  0.87581   & -0.00085    &  0.25000  & 8$g$   &  \\
 & Ti  &  0.12418   &  0.25082    &  0.25000  & 8$g$   &  \\
 & Sb  &  0.00000   &  0.37533    &  0.12909  & 8$f$   &  \\
 & Sb  &  0.00000   &  0.12532    &  0.62908  & 8$f$   &  \\
 & Sb  &  0.75033   &  0.12499    &  0.12907  & 16$h$   &  \\
 &  O  &  0.00000   &  0.62500    &  0.25000  & 4$c$   &  \\
 &  O  &  0.00000   &  0.12500    &  0.25000  & 4$c$   &  \\
 &  O  &  0.75000   &  0.37500    &  0.25000  & 8$g$   &  \\
\hline
\end{tabular}
\end{center}
\end{table}
%%%%%%%%%%%%%%%%%%

%%%%%%%%%%%%%%%%%%%%%%%%%%%%%%%
\bibliographystyle{apsrev4-1}
\bibliography{references}

%%%%%%%%%%%%%%%%%%%%%%%%%%%%%%%

\end{document}